\documentclass[12pt]{iopart}
\usepackage{epsfig}

\newcommand{\braket}[2]{\langle #1 \,|\, #2 \rangle}
\newcommand{\ket}[1]{| \, #1 \rangle}
\newcommand{\bra}[1]{ \langle #1 \,  |}

\begin{document}
\title{A necessary and sufficient condition to play games in quantum mechanical settings}
\author{S K \"Ozdemir, J Shimamura and N Imoto}
\address{SORST Research Team for Interacting Carrier
Electronics,\\
CREST Research Team for Photonic Quantum Information,\\ Graduate
School of Engineering Science, Osaka University, 1-3
Machikaneyama, Toyonaka, Osaka 560-8531, Japan}
\ead{ozdemir@qi.mp.es.osaka-u.ac.jp}

\pagestyle{plain} \pagenumbering{arabic}

\begin{abstract}Quantum game theory is a multidisciplinary field which
combines quantum mechanics with game theory by introducing
non-classical resources such as entanglement, quantum operations
and quantum measurement. By transferring two-player-two strategy
($2\times 2$) dilemma containing classical games into quantum
realm, dilemmas can be resolved in quantum pure strategies if
entanglement is distributed between the players who use quantum
operations. Moreover, players receive the highest sum of payoffs
available in the game, which are otherwise impossible in classical
pure strategies. Encouraged by the observation of rich dynamics of
physical systems with many interacting parties and the power of
entanglement in quantum versions of $2\times 2$ games, it became
generally accepted that quantum versions can be easily extended to
$N$-player situations by simply allowing $N$-partite entangled
states. In this article, however, we show that this is not
generally true because the reproducibility of classical tasks in
quantum domain imposes limitations on the type of entanglement and
quantum operators. We propose a benchmark for the evaluation of
quantum and classical versions of games, and derive the necessary
and sufficient conditions for a physical realization. We give
examples of entangled states that can and cannot be used, and the
characteristics of quantum operators used as strategies.
\end{abstract}

\section{Introduction} Mathematical models and techniques of game theory have
increasingly been used by computer and information scientists,
i.e., distributed computing, cryptography, watermarking and
information hiding tasks can be modelled as games
\cite{Moulin,Cohen,Conway,Shen,Ettinger,Pateux}. Therefore, new
directions have been opened in the interpretation and use of game
theoretical toolbox which has been traditionally limited to
economical and evolutionary biology problems \cite{Books}. This is
not a surprise because all have information as the common
ingredient and the strong connection among them \cite{Lee1}: Game
theory deals with situations where players make decisions which
affect the outcomes and payoffs. All the involved processes can be
modelled as information flow. Since physical systems, which are
governed by the laws of quantum mechanics, are used during
information flow (generation, transmission, storage and
manipulation), game theory becomes closely related to quantum
mechanics, physics, computation and information sciences. Along
this line of thinking, researchers introduced the quantum
mechanical toolbox into game theory to see what new features will
arise combining these two beautiful areas of science
\cite{Eisert1,Meyer,Weber,Ben,Du2,Flitney,Shima,Ozdemir,Lee}.

Quantum mechanics is introduced into game theory through the use
of quantum bits (qubits) instead of classical bits, quantum
operations and entanglement which is a quantum correlation with a
highly complex structure and is considered to be the essential
ingredient to exploit the potential power of quantum information
processing. This effort, although has been criticized on the basis
of using artificial models \cite {Ben1,Enk}, has produced
significant results: (i) Dilemmas in some games can be resolved
\cite{Eisert,Eisert1,Ozdemir,Shima,Simon,Adrian}, (ii) playing
quantum games can be more efficient in terms of communication
cost; less information needs to be exchanged in order to play the
quantized versions of classical games \cite{Lee,Lee1,Ozdemir},
(iii) entanglement is not necessary for the emergence of Nash
Equilibrium but for obtaining the highest possible sum of payoffs
\cite{Ozdemir}, and (iv) quantum advantage does not survive in the
presence of noise above a critical level \cite{Johnson3,Ozdemir1}.
In addition, market phenomena, bargaining, auction and finance
have been described using quantum game theory \cite{Edward1}. The
positive results are consequences of the fact that quantum
mechanical toolbox allows players to have a larger set of
strategies to choose from when compared to classical games.

In this paper, we focus on the extent of entangled states and
quantum operators that can and cannot be used in multi-player
games, and introduce a benchmark for the comparison of classical
games and their quantized versions on a fair basis. Moreover, this
study attempts to clarify a relatively unexplored area of interest
in quantum game theory, that is the effect of different types of
entangled states and their use in multi-player multi-strategy
games in quantum settings. Our approach is based on the
reproducibility of classical games in the physical schemes used
for the implementation of their quantized versions.

Reproducibility requires that a chosen model of game should
simulate both quantum and classical versions of the game to allow
a comparative analysis of quantum and classical strategies, and to
discuss what can or cannot be attained by introducing quantum
mechanical toolbox. This is indeed what has been observed in
quantum Turing machine which can simulate and reproduce the
results of the classical Turing machine. Therefore, the
reproducibility criterion must be taken into consideration
whenever a comparison between classical and quantum versions of a
task is needed. An important consequence of this criterion in game
theory is the main contribution of this study: Derivation of the
necessary and sufficient condition for entangled states and
quantum operators that can be used in the quantized versions of
classical games.

\section{Multiplayer games}
\subsection{ Definitions and model}
In classical game theory, a strategic game is defined by
$\Gamma=[N,(S_{i})_{i\in N},(\$_{i})_{i\in N}]$ where $N$ is the
set of players, $S_{i}$ is the set of pure strategies available to
the $i$-th player, and $\$_{i}$ is his payoff function from the
set of all possible pure strategy combinations ${\cal
C}=\times_{j\in N}S_{j}$ into the set of real numbers ${\bf R}$.
When the strategic game $\Gamma$ is played with pure strategies,
each player $i$ choose only one of the strategies $s_i$ from the
set $S_i$. With each player having $m$ pure strategies, $\cal{C}$
has $m^N$ elements. Then for the $k$-th joint strategy
$c_k\in\cal{C}$,  payoffs of each player can be represented by an
ordered vector $A_k=(a_{k}^{1},a_{k}^{2},...,a_{k}^{N})$ where
$a_{k}^{j}=\$_{j}(c_k)$ is the payoff of the $j$-th player for the
$k$-th joint strategy outcome. Players may choose to play with
mixed strategies (randomizing among pure strategies) resulting in
the expected payoff
\begin{eqnarray}\label{mix:1} f_i(\overline{q}_1,\cdot\cdot,\overline{q}_N) =
\sum_{c_k \in \cal{C}}\left(\prod_{j\in N} q_j (s_j)\right)
\$_i(c_k)= \sum_{k=1}^{m^N}\left(\prod_{j=1}^{N} q_j (s_j)\right)
a_k^i
\end{eqnarray}
where $f_i(\overline{q}_1,\cdot\cdot,\overline{q}_N)$ is the
payoff of the $i$-th player for the probability distributions
$\overline{q}_{t}$ over the strategy set $S_{t}$ of each player
$t$, and $q_j(s_j)$ represents the probability that $j$-th player
chooses the pure strategy $s_j$ according to the distribution
$\overline{q}_{j}$.

Most of the studies on quantum versions of classical games have
been based on the model proposed by Eisert {\it et al.}
\cite{Eisert1}. In this model, the strategy set of the players
consists of unitary operators which are applied locally on a
shared entangled state. A measurement by a referee on the final
state after the application of the operators maps the chosen
strategies of the players to their payoffs. For example, the
strategies ``Cooperate" and ``Defect" of players in classical
Prisoner's Dilemma is represented by the unitary operators,
$\hat{\sigma}_0$ and $i\hat{\sigma}_{y}$.

In this study, however, we consider a more general model than
Eisert {\it et al.}'s model \cite{Eisert1} for
$N$-player-two-strategy games. In our model \cite{Shima1}, (i) A
referee prepares an $N$-qubit entangled state $\ket{\Psi}$ and
distributes it among $N$ players, one qubit for each player. (ii)
Each player independently and locally applies an operator chosen
from the entire set of special unitary operators for dimension
two, SU(2), on his qubit. Assuming that $i$-th player applies
$\hat{u}_i$,  the joint strategy of all the players is represented
by the tensor product of unitary operators as $\hat{x}=\hat{u}_1
\otimes \hat{u}_2 \otimes \cdots \otimes \hat{u}_N$, which
generates the output state $\hat{x}\ket{\Psi}$ to be submitted to
the referee. (iii) Upon receiving this final state, the referee
makes a projective measurement $\{\hat{\mathcal{P}}_j
\}_{j=1}^{2^{N}}$ which outputs $j$ with probability ${\rm
Tr}[\hat{\mathcal{P}}_{j}
\hat{x}\ket{\Psi}\bra{\Psi}\hat{x}^{\dagger}]$, and assigns
payoffs chosen from the payoff matrix depending on the measurement
outcome $j$. Therefore, the expected payoff of the $i$-th player
is described by
\begin{eqnarray}
f_{i}(\hat{x})={\rm Tr}\left[\left(\Sigma_j
a_{j}^i\hat{\mathcal{P}}_{j}\right)\hat{x} \ket{\Psi}
\bra{\Psi}\hat{x}^{\dagger}\right]\label{eq:quantum}
\end{eqnarray}
where $\hat{\mathcal{P}}_j$ is the projector and $a_j^{i}$ is the
$i$-th player's payoff when the measurement outcome is $j$.  This
model can be implemented in a physical scheme with the current
level of quantum technology.

\subsection{Classification of $N$-player two-strategy games:}
In general, one can prepare a large number of generic games by
arbitrarily choosing the entries of game payoff matrix. However,
not all of those generic games are interesting enough to be the
subject of game theory. Classical game theory mainly focuses on
specific dilemma containing $2\times 2$ games such as Prisoner's
dilemma (PD), Stag-Hunt (SH), Chicken Game (CG), Dead-Lock (DL),
Battle of Sexes (BoS) Samaritan's dilemma (SD), Boxed Pigs (BP),
Modeller's dilemma (MD), Ranked Coordination (RC), Alphonse \&
Gaston Coordination Game (AG), Hawk-Dove (HD), Battle of Bismarck
(BB), Matching Pennies (MP) \cite{Books}. Multi-player extensions
of these $2\times 2$ games and some originally multi-player games,
such as minority and coordination games which have direct
consequences where populations are forced to coordinate their
actions, are also the subject of game theory. In this study, we
consider only those interesting games instead of studying all
generic games that can be formed.

In an $N$-player game, every player plays one of his strategies
against all other $N-1$ players simultaneously. The payoff matrix
of an $N$-player two-strategy game is characterized by $2^N$
possible outcomes and a total of $N2^N$ parameters. Payoffs of
each player for the $k$-th possible outcome can be represented by
an ordered vector $A_k=(a_{k}^{1},a_{k}^{2},...,a_{k}^{N})$. Based
on the payoffs for all possible outcomes, we group the games into
two: {\it Group I} contains the games where all the outcomes have
different payoff vectors, that is $A_j\neq A_k$ for $\forall k
\neq j$, whereas {\it Group II} contains the games where payoff
vectors for some outcomes are the same, $A_j=A_k$, implying
$(a_{j}^{1},a_{j}^{2},...,a_{j}^{N})=(a_{k}^{1},a_{k}^{2},...,a_{k}^{N})$,
for $\exists k \neq j$.

When a two-player two-strategy game is extended to $N$-player game
($N>2$), the new payoff matrix is formed by summing the payoffs
that each player would have received in simultaneously playing the
two-player game with $N-1$ players. Hence, in their $N$-player
extensions, the games PD, SD, BP, MD, DL, and RC fall into {\it
Group I} while BoS, BB, MP, and AG games in {\it Group II}. For
$N=3$, BoS becomes a member of {\it Group I}.  CG, SH and HD
belong to either the first or second group according to whether
$N$ is even or odd. For even $N$, SH belongs to {\it Group I} and
CG and HD belong to {\it Group II}, and vice verse. Minority,
majority and coordination Games are in {\it Group II}.

\section{Reproducibility criterion to play games in quantum mechanical settings} We consider the
reproducibility of a multi-player two-strategy classical game in
the quantization model explained above. First, reproducibility
problem in pure strategies will be discussed in details, and later
the conditions for mixed strategies will be given. We require that
a classical game be reproduced when each player's strategy set is
restricted to two unitary operators, $\{
\hat{u}_{i}^{1},\hat{u}_{i}^{2} \}$, corresponding to the two pure
strategies in the classical game. Then the joint pure strategy of
the players is represented by $\hat{x}_k=\hat{u}_{1}^{l_1} \otimes
\hat{u}_{2}^{l_2} \otimes \cdots \otimes \hat{u}_{N}^{l_N}$ with
$l_i=\{1,2\}$ and $k = \{1,2, \ldots, 2^N \}$. Thus the output
state becomes $\ket{\Phi_k}=\hat{x}_k\ket{\Psi}$. For the strategy
combination $\hat{x}_{k}$, expected payoff for the $i$-th player
becomes as in Eq. (\ref{eq:quantum}) with $\hat{x}$ replaced by
$\hat{x}_k$. Then $A_k$ defined in the previous section is the
ordered payoff vector of all players for the $k$-th possible
outcome.

Reproducibility problem can be stated in two cases: In {\it CASE
I}, the referee should be able to identify the strategy played by
each player deterministically regardless of the structure of the
payoff matrix, whereas in {\it CASE II} the referee should be able
to reproduce the expected payoff given in eq. (\ref{mix:1}) in the
quantum version, too, \cite{Shima1}.  While in {\it CASE I} the
referee needs to identify all possible outcomes, in {\it CASE II}
he just needs to distinguish between the sets of outcomes with the
same payoff. {\it CASE II} is equivalent to {\it CASE I} for {\it
Group I} games where all outcomes of the game have different
payoff vectors. We call the situations described in {\it CASE I}
and its equivalence in {\it CASE II} as the ''strong criterion,"
and the rest of the situations as the ''weak criterion" of
reproducibility.

\subsection{The strong criterion of reproducibility (SCR)}
This criterion requires that referee discriminate all the possible
output states $\ket{\Phi_k}$ deterministically in order to assign
payoffs uniquely in the pure strategies. That is, the projector
$\{ \hat{\mathcal{P}}_j \}_{j=1}^{2^{N}}$ has to satisfy ${\rm
Tr}[\hat{\mathcal{P}}_{j} \ket{\Phi_k}\bra{\Phi_k}]=\delta_{jk}$,
which is possible if and only if
\begin{equation}
\braket{\Phi_\alpha}{\Phi_\beta}=\delta_{\alpha\beta} \;\; \forall
\alpha, \beta. \label{condition}
\end{equation}
Thus, SCR transforms the reproducibility problem into quantum
state discrimination where we know that two quantum states can be
deterministically discriminated iff they are orthogonal. Under
SCR, we see that
$f_i(\overline{q}_1,\cdot\cdot,\overline{q}_N)=f_i(\hat{x}_k)=a_k^i$
because there is no randomization over the strategy sets (each
player choose one and only one strategy deterministically) and the
only outcome is $\hat{x}_k$ with probability one. Therefore, Eq.
(\ref{condition}) becomes the {\it necessary condition} for the
strong reproducibility criterion (SCR).

Among the multi-partite ($N\geq 3$) entangled states we focused on
GHZ-like states of the form $\ket{{\rm GHZ}}_N=(\ket{00 \ldots 0}
+ i \ket{11 \ldots 1})/\sqrt{2}$ and symmetric Dicke states
represented as $\ket{N-m,m}/\sqrt{{}_{N}C_m}$ with $(N-m)$ zeros
and $m$ ones (${}_{N}C_m$ denoting the binomial coefficient).
Imposing SCR we observed \cite{Shima1} the following.
\begin{description}
\item[(a)]For Dicke states with unequal number of zeros and ones
($N$-party W-state, defined as $|W_{N}\rangle=|N-1,1\rangle/
{\sqrt{N}}$ is a member of this class),
\begin{description}
\item[({\it a1})] $\hat{u}_k^{1\dagger} \hat{u}_k^2
=\hat{\sigma_x} \hat{R}_z(2\phi_k)$ for any two output states
differing only in $k$-th player's strategy where the rotation
operator $\hat{R}_z(\gamma)$ is defined as
$\hat{R}_z(\gamma)=e^{-i\gamma\hat{\sigma}_z/2}$, and

\item[({\it a2})] $\phi_j -\phi_k=n\pi+\pi/2$ for any two output
states different only in the strategies of $j$-th and $k$-th
players.
\end{description}

Then for any three players $j,k,m$ participating the game, we
obtain the set of equations $\chi_{jkm}=\{\phi_j
-\phi_k=n\pi+\pi/2$, $\phi_m -\phi_j=n'\pi+\pi/2$, $\phi_k
-\phi_m=n''\pi+\pi/2\}$ where $n$, $n'$ and $n''$ are integer. The
sum of the three equations in $\chi_{jkm}$ results in
$3\pi/2+m'\pi=0$ which is satisfied for $m'=-3/2$; however this
contradicts the fact $m'=n+n'+n''$ is an integer. \item[(b)] For
Dicke states $|N/2,N/2\rangle$ with even $N\geq 6$,
\begin{description}
\item[({\it b1})] $\hat{u}_k^{1\dagger} \hat{u}_k^2
=\cos\theta_k\hat{\sigma}_z+\sin\theta_k\hat{\sigma_x}
\hat{R}_z(2\phi_k)$ with real $\theta_k$ and $\phi_k$ is a
two-parameter SU(2) operator obtained from the mutual
orthogonality of two output states which differ in the strategies
of one player,

\item[({\it b2})]
$\cos\theta_k\cos\theta_j=(N/2)\cos(\theta_k-\theta_j)\sin\theta_k\sin\theta_j$
from the inner product of two states which differ only in the
strategies of two-players, and

\item[({\it b3})] from the output states which differ in the
strategies of four players $i,j,k,l$,
\begin{eqnarray}\label{N13}
&&\frac{24}{N(N-2)}\cos{\theta_{i}}\cos{\theta_{j}}\cos{\theta_{k}}\cos{\theta_{l}}\nonumber\\
&&=[\cos\beta_{1}+\cos\beta_{2}+\cos\beta_{3}]\sin{\theta_{i}}\sin{\theta_{j}}\sin{\theta_{k}}\sin{\theta_{l}}
\end{eqnarray}
where $\beta_{1}=\phi_{i}+\phi_{j}-\phi_{k}-\phi_{l}$,
$\beta_{2}=\phi_{i}-\phi_{j}+\phi_{k}-\phi_{l}$ and
$\beta_{3}=\phi_{i}-\phi_{j}-\phi_{k}+\phi_{l}$.
\end{description}

Then we obtain $\theta_i\neq n\pi$ and $\theta_i\neq \pi/2+n\pi$
for $\forall i$ using ({\it b1,b2}) and ({\it b1,b2,b3}),
respectively. Next, we write ({\it b2}) for the pair of players
$(i,j)$ and $(k,l)$ and multiply these two equations. Doing the
same for different pairs of players $(i,k)$ and $(j,l)$, and
comparing the final expressions with Eq. \ref{N13}, we find
$\theta_i=\pi/2+n\pi$ for $\forall i$ which contradicts the above
result obtained from ({\it b1,b2,b3}).
\end{description}

If the mutual orthogonality relations lead to {\it contradictions}
outlined in (a) and (b), the corresponding entangled state cannot
be used in quantum versions of classical games under SCR. Among
the class of entangled states studied we have found: (i) bell
states and any two-qubit pure state satisfy SCR if the unitary
operators for the players are chosen as
$\{\hat{\sigma}_0,\hat{\sigma}_{x} \}$ and $\{ \hat{\sigma}_0,
i\hat{\sigma}_{y} \}$. (ii) $\ket{{\rm GHZ}}_N$ satisfies SCR if
the unitary operators of the players are chosen as $\{
\hat{\sigma}_0, i\hat{\sigma}_y \}$. Entangled states that can be
obtained from $\ket{{\rm GHZ}}_N$ state by local unitary
transformations also satisfy SCR. (iii) $|W_{N}\rangle$ does not
satisfy SCR, therefore cannot be used in this model of quantum
games. (d) Among the Dicke states, only the states $\ket{1,1}$ and
$\ket{2,2}$ satisfy the SCR. These results are valid for all the
games in {\it Group I} and the situations where {\it CASE I} is
desired.
\subsubsection{Quantum operators and SCR}
Assume that there are two unitary operators corresponding to the
classical pure strategies for the entangled state, $\ket{\Psi}$.
Imposing SCR on the situation where two outcomes, $\ket{\Phi_0}=
\hat{u}_1^1 \otimes \hat{u}_2^1 \otimes \cdots \otimes
\hat{u}_N^1\ket{\Psi}$ and $\ket{\Phi_1}= \hat{u}_1^2 \otimes
\hat{u}_2^1 \otimes \cdots \otimes \hat{u}_N^1\ket{\Psi}$, differ
only in the operator of the first player, we find
\begin{equation}
\bra{\Psi} (\hat{u}_1^1)^{\dagger}\hat{u}_1^2 \otimes \hat{I}
\otimes \cdots \otimes \hat{I} \ket{\Psi} = 0. \label{eq1:con}
\end{equation}
Since $(\hat{u}_1^1)^{\dagger}\hat{u}_1^2$ is a normal operator,
it can be diagonalized by a unitary operator $\hat{z}_1$.
Furthermore, since $(\hat{u}_1^1)^{\dagger}\hat{u}_1^2$ is a SU(2)
operator, the eigenvalues are given by $e^{i \phi_ 1}$ and $e^{-i
\phi_1}$. Then Eq. (\ref{eq1:con}) can be transformed into
\begin{eqnarray}
&&\hspace{-15mm}\bra{\Psi}\, \hat{z}_1^{\dagger} \,
(\hat{z}_1(\hat{u}_1^1)^{\dagger}\hat{u}_1^2 \hat{z}_1^{\dagger})
\, \hat{z}_1 \otimes
\hat{I} \otimes \cdots \otimes \hat{I} \,\ket{\Psi} \nonumber \\
&&\hspace{35mm} =  \bra{\Psi'} R_z(-2\phi_1)\otimes \hat{I}
\otimes \cdots \otimes \hat{I} \ket{\Psi '} \nonumber \\
&&\hspace{35mm} = \cos \phi_1 + i \left(2\Sigma_{i_j \in \{0,1\}}
|c_{0 \, i_2 ... i_N}|^2 - 1\right)\sin \phi_1 = 0 \label{eq2:con}
\end{eqnarray} where $\ket{\Psi'}= \hat{z}_1 \otimes \hat{I} \otimes \cdots
\otimes \hat{I} \ket{\Psi}$  is written on computational basis as
$\ket{\Psi'} = \Sigma_{i_j \in \{0,1\}} c_{i_1 \, i_2 ... i_N}
\ket{i_1}\ket{i_2}\cdots\ket{i_N}$.

In order for the above equality to hold, $\cos\phi_1 = 0$ and
$2\Sigma_{i_j \in \{0,1\}} |c_{0 \, i_2 ... i_N}|^2 - 1=0$ must be
satisfied. The equation $\cos\phi_1 = 0$ implies that the
diagonalized form $\hat{D}_1 = \hat{z}_1(
\hat{u}_1^1)^{\dagger}\hat{u}_1^2
\hat{z}_1^{\dagger}=i\hat{\sigma}_z$. This argument holds for all
players, therefore we write $\hat{z}_k
(\hat{u}_k^1)^{\dagger}\hat{u}_k^2 \hat{z}_k^{\dagger} =\hat{D}_k
=i\hat{\sigma}_z$ for $k=1,\cdots,N$. For example, in the case of
the Dicke state $\ket{2,2}$, which satisfies SCR with the unitary
operators $\hat{u}_{k=1,2,3,4}^1=\hat{I}$,
$\hat{u}_{k=1,2,3}^2=i(\sqrt{2}\hat{\sigma}_{z}+\hat{\sigma}_{x})/\sqrt{3}$
and $\hat{u}_{4}^2=i\hat{\sigma}_{y}$, it is easy to verify that
eigenvalues $\mp i$ of $\hat{u_k}^{1\dagger}\hat{u}_k^2$ are
already in the diagonalized form. For GHZ state, the operators are
$\hat{u}_{k}^1=\hat{I}$ and $\hat{u}_{k}^2=i\hat{\sigma}_y$ which
can be written in the form $\hat{D}_1$.

 Next we consider the following scenario: Each player has
two operators satisfying the above properties. Instead of choosing
either of these operators, they prefer to use a linear combination
of their operator set. Let this operator be $\hat{w}_k =
\hat{u}_k^1\cos \theta_k  + \hat{u}_k^2\sin \theta_k $ for the
$k$-th player. Then, we ask  (i) Does the property of the
operators $\hat{u}_k^1$ and $\hat{u}_k^2$ derived from the SCR
impose any condition on the operator $\hat{w}_k$?, and (ii) What
does the outcome of the game played in the quantum version with
the operator $\hat{w}_k$ imply? Since $\hat{z}_k
(\hat{u}_k^1)^{\dagger}\hat{u}_k^2\hat{z}_k^{\dagger}$ is in the
diagonalized form we can write
\begin{eqnarray}\label{eq1970:unitw}
\hat{w}_k^{\dagger} \hat{w}_k =\hat{I} + \cos\theta_k \sin\theta_k
(\hat{z}_k^{\dagger} \hat{D}_k\hat{z}_k +\hat{z}_k^{\dagger}
\hat{D}_k^{\dagger} \hat{z}_k  )= \hat{I},
\end{eqnarray} where we have used $\hat{u}_k^{1\dagger}\hat{u}_k^2 =
\hat{z}_k^{\dagger} \hat{D}_k\hat{z}_k$, and $\hat{D}^{\dagger}_k=
- \hat{D}_k$ since $\hat{D}_k$ is anti-hermitian. This equation
implies that SCR requires $\hat{w}_k$ be a unitary operator.

When the players use the operators $\hat{w}_k=\hat{u}_k^1\cos
\theta_k + \hat{u}_k^2\sin \theta_k $, the joint strategy
$\hat{x}$ becomes
$\hat{x}=\hat{w}_1\otimes\hat{w}_2\otimes\ldots\otimes\hat{w}_N=\bigotimes_{j\in
N}\hat{w}_j$. Substituting $\hat{x}$ into Eq. (\ref{eq:quantum}),
we obtain
\begin{eqnarray}
f_i(\hat{x})=\sum_{\mu=1}^{2^N}\left(\prod_{\ell=1}^{\mu-1}\sin^2\theta_{\ell}\right)
\left(\prod_{j=\mu}^{N}\cos^2\theta_{j}\right)a_{\mu}^{i}.
\label{eq:reproduce}
\end{eqnarray} Note that Eq. (\ref{eq:reproduce}) has the same form of Eq.
(\ref{mix:1}) implying that payoffs of the classical mixed
strategies are reproduced in the quantum version for $\hat{w}_k $.
Therefore, we conclude that Eq. (\ref{condition}) is the {\it
necessary and sufficient condition} for the reproducibility of a
classical game in the quantum version according to SCR. This is
because when players apply one of their pure strategies
$\hat{u}_k^{1}$ or $\hat{u}_k^2$ with unit probability, results of
classical pure strategy; when they apply a linear combination of
their pure strategies results of classical mixed strategy are
reproduced in the quantum setting. Another way of reproducing the
results of classical mixed strategies is that players apply their
pure strategies $\hat{u}_k^{1}$ and $\hat{u}_k^2$ according to a
probability distribution as is the case in classical mixed
strategies. Note that this is different than applying a linear
combination of their pure strategies $\hat{u}_k^{1}$ and
$\hat{u}_k^2$.
\subsubsection{Entangled states and SCR} After stating the properties of
operators which satisfy SCR, we proceed to investigate the
properties of the class of entangled states which satisfy it.
Suppose that an N-qubit state $\ket{\Psi}$ and two unitary
operators $\{\hat{u}_k^1, \hat{u}_k^2\}$ satisfy SCR. Then for two
possible outcomes $\ket{\Phi_0}= \hat{u}_1^1 \otimes \hat{u}_2^1
\otimes \cdots \otimes \hat{u}_N^1 \ket{\Psi}$ and $\ket{\Phi_1}=
\hat{u}_1^2 \otimes \hat{u}_2^1 \otimes \cdots \otimes
\hat{u}_N^1\ket{\Psi}$, Eq. (\ref{condition}) requires
\begin{eqnarray}
\bra{\Psi} \hat{z}_1^{\dagger} \, \hat{z}_1\hat{u}_k^{1\dagger}
\hat{u}_k^2 \hat{z}_1^{\dagger} \, \hat{z}_1\otimes \hat{I}
\otimes \cdots \otimes \hat{I} \ket{\Psi}=\bra{\Psi'}\hat{D}_1
\otimes \hat{I} \otimes \cdots \otimes \hat{I} \ket{\Psi'} =0,
\label{eq3:con}
\end{eqnarray}
where $\hat{z}_1$ is a unitary operator diagonalizing
$\hat{u}_1^{1\dagger} \hat{u}_1^2$ and $\ket{\Psi'} = \hat{z}_1
\otimes \hat{I} \otimes \cdots \otimes \hat{I} \ket{\Psi}$. This
implies that if the N-qubit state $\ket{\Psi}$ and the operators
$\{\hat{u}_k^1, \hat{u}_k^2\}$ satisfy Eq. (\ref{condition}), then
the state $\ket{\Psi '} = \hat{z}_1 \otimes \hat{z}_2 \otimes
\cdots \otimes \hat{z}_N \ket{\Psi}$ and the unitary operators
$\{\hat{D}, \hat{I} \}$ should satisfy, too. Since the global
phase is irrelevant, Eq. (\ref{eq3:con}) can be further reduced to
$\bra{\Psi'}\hat{\sigma}_{z} \otimes \hat{I} \otimes \cdots
\otimes\hat{I} \ket{\Psi'} = 0$. Thus, we end up with $2^{N}-1$
equalities to be satisfied:
\begin{eqnarray}
&&\bra{\Psi'} \hat{\sigma}_z \otimes \hat{I}\otimes \hat{I}
\otimes \cdots
\otimes \hat{I} \ket{\Psi'} = 0, \nonumber \\
&& \bra{\Psi'} \hat{I} \otimes \hat{\sigma}_z \otimes \hat{I}
\otimes \cdots
\otimes \hat{I} \ket{\Psi'} = 0, \nonumber \\
&&  \;\;\;\;\;\;\;\;\;\; \vdots \nonumber \\
&&\bra{\Psi'} \hat{\sigma}_z \otimes \hat{\sigma}_z \otimes \cdots
\otimes \hat{\sigma}_z \ket{\Psi'} = 0.
\end{eqnarray}\label{eq10}Defining $\ket{\Psi'}=\Sigma_{i_j \in \{ 0,1 \}} c_{i_1 i_2 ...
i_N} \ket{i_1}\ket{i_2}\cdots\ket{i_N}$, we write Eq.
(\ref{eq3:con}) in the matrix form as
\begin{equation}
\left[
\begin{array}{ccccc}
1 & 1 & \ldots & -1 & -1 \\
  &  & \ldots &   &     \\
  &   &  \vdots    &   &    \\
1 & 1 & \ldots & 1 & 1
\end{array}
\right] \left[
\begin{array}{l}
|c_{00...0}|^2 \\
|c_{00...1}|^2 \\
  \;\;\;\; \vdots \\
|c_{11...1}|^2 \\
\end{array}
\right] = \left[
\begin{array}{l}
0 \\
0 \\
\vdots \\
1 \\
\end{array}
\right], \label{eq:diagonal}
\end{equation}
where the last row is the normalization condition. The row vector
corresponds to the diagonal elements of $\hat{\sigma}_z^{\{0,1\}}
\otimes \cdots \otimes \hat{\sigma}_z^{\{0,1\}}$ where
$\hat{\sigma}_z^{0}$ is defined as $\hat{I}$. Consider the
operators $\hat{x},\hat{y} \in (\hat{\sigma}_z^{\{0,1\}})^{\otimes
N}$ where $\hat{x}\hat{y} \in (\hat{\sigma}_z^{\{0,1\}})^{\otimes
N}$. Since ${\rm Tr}[\hat{\sigma}_z] = 0$, for $\hat{x} \ne
\hat{y}$, we have ${\rm Tr}[\hat{x}\,\hat{y}] = {\rm
Tr}[\hat{x}]\,{\rm Tr}[\hat{y}] = 0$. Thus any two row vectors are
orthogonal to each other, thus the matrix in Eq.
(\ref{eq:diagonal}) has an inverse, and $|c_{i_1 i_2 ... i_N}|^2$
are uniquely determined as $1/N$. This implies that if a state
satisfies SCR, then it should be transformed by local unitary
operators into the state which contains all possible terms with
the same magnitude but different relative phases:
\begin{equation} \ket{\Psi'} =
\frac{1}{\sqrt{N}}\sum_{i_j \in \{ 0,1 \}}
 e^{i \phi_{i_1 i_2 .. i_N}}
\ket{i_1}\ket{i_2}\cdots\ket{i_N}.\label{eq15:st}
\end{equation}
One can show that product state and GHZ state, which satisfy SCR,
can be transformed into the form of Eq. (\ref{eq15:st}),
respectively, by Hadamard operator, $\hat{H}=(\hat{\sigma_x} +
\hat{\sigma_z})/\sqrt{2}$, and by $(e^{i \frac{\pi}{4}} \hat{I} +
e^{- i \frac{\pi} {4}} \hat{ \sigma}_z + \hat{\sigma}_y
)/\sqrt{2}$ for one player and $\hat{H}$ for the others.

\subsection{The weak criterion of reproducibility (WCR)}
This weak version of the reproducibility criterion requires that
referee deterministically discriminate all the possible sets
formed by the output states with the same payoff vectors in order
to assign payoffs uniquely in the pure strategies. When $A_j=A_k$,
output states $\ket{\Phi_j}$ and $\ket{\Phi_k}$ should be grouped
into the same set. If all possible output states are grouped into
sets
$S_j=\{|\Phi_{1j}\rangle,|\Phi_{2j}\rangle,\dots|\Phi_{nj}\rangle\}$
and
$S_k=\{|\Phi_{1k}\rangle,|\Phi_{2k}\rangle,\dots|\Phi_{n'k}\rangle\}$
then the referee should deterministically discriminate between
these sets which is possible iff the state space spanned by the
elements of each set are orthogonal. Hence for every element of
$S_j$ and $S_k$, we have $\langle\Phi_{nj}|\Phi_{n'k}\rangle=0$,
$\forall j\neq k$, that is all the elements of $S_j$ and $S_k$
must be orthogonal to each other, too. Thus WCR transforms the
reproducibility problem into set discrimination problem. We named
it as WCR because the condition of sets $S_1,..,S_k$ being
mutually orthogonal to each other is a much looser condition than
the condition of all states in $S=\bigoplus_i^{k} S_i$ being
mutually orthogonal to each other. The sets $S_1,..,S_k$ may be
mutually orthogonal even if the states in $S$ are linearly
dependent. If we relax the criterion of deterministic
discrimination and allow inconclusive results then one can use
unambiguous state and set discrimination. However, we are not
concerned with this situation because we require that classical
game is reproduced in the quantum settings deterministically. It
is clear that the games in {\it Group II} should be discussed with
WCR. A natural question is whether the results listed in (a)-(d)
are valid for {\it Group II} games or not. The answer to this
question will be given below.

\subsubsection{Entangled states and WCR} In this section we check
whether the results obtained under SCR is valid or not for {\it
Group II} games with WCR. We start by asking the question ``{\it
Is there a partition of all possible outcomes (output states) into
sets such that mutual orthogonality of these sets does not lead to
the contradictions discussed for SCR?}"  The following
observations from the analysis of SCR for a given entangled state
makes our task easier:

({\it O1}) For $\ket{N-m,m}$ with $N\neq m$, if the mutual
orthogonality condition of the sets leads to the operator form as
in ({\it a1}) for all players, then there will be {\it
contradiction} if we obtain the set $\chi_{jkm}$ for any
three-player-combination $(j,k,m)$. Presence of at least one such
set is enough to conclude that there is {\it contradiction}. On
the other hand, to prove that there is no contradiction, one has
to show that at least one of the equations in $\chi_{jkm}$ is
missing for all three-player-combinations.

({\it O2}) For $\ket{N-m,m}$ with $N\neq m$, if the mutual
orthogonality condition of the sets leads to the operator form as
in ({\it a1}) for one and only one player, then there will be {\it
no contradiction} because there will be at least one missing
equation in $\chi_{jkm}$ for all possible
three-player-combinations $(j,k,m)$. Note that such a situation
occurs iff $2^N$ possible outputs are divided into two sets with
equal number of elements. Then the only equations we will obtain
are $\phi_1 -\phi_j=n\pi+\pi/2$ for all $j={2,\cdots,N}$.

({\it O3}) For $\ket{N/2,N/2}$ with $N\geq 6$, a necessary
condition for contradiction is to have equations of the form ({\it
b2}) for at least four different pairing of players, such as
$\{(i,j),(k,l)\}$ and $\{(i,k),(j,l)\}$. If the mutual
orthogonality condition of the sets leads to the operator form as
in ({\it b1}) for one and only one player, say first player, then
from ({\it b2}) we will obtain only
$\cos\theta_1\cos\theta_j\exp(\varphi_j)=(N/2)\cos(\theta_1-\theta_j)\sin\theta_1\sin\theta_j$
for all $j={2,\cdots,N}$ where $\exp(\mp\varphi_j)$ denotes the
phase of the diagonal elements of the matrix
$u^{1\dagger}_ju^{2}_j$. This extra phase parameter and the
absence of similar relations between players other than the first
allow us to freely set the operator parameters for all players.
Therefore,  {\it no contradiction} occurs.

({\it O4}) For $\ket{N/2,N/2}$ with $N\geq 4$, if the mutual
orthogonality of states leads to the relations in ({\it b1}) and
({\it b2}) then contradiction will not occur iff the outcomes
differing in the strategies of four players are in the same set.

({\it O5}) For all $\ket{N-m,m}$ except $\ket{1,1}$ and
$\ket{2,2}$, if the number of elements in any of the sets in a
{\it Group II} game is an odd number, then there will always be
{\it contradiction}. If one of the output states in any set is
left alone then this state will satisfy the mutual orthogonality
condition with the elements of the other sets which will lead to
the relations mentioned above, and hence to {\it contradiction}.

({\it O6}) If there is a set with only two elements which are the
outcomes when all the players choose the same strategy, there will
be {\it contradiction}.

Our analysis revealed that multiparty extensions of $2\times2$
games have payoff structures such that partitioning results in one
or more sets with only one element. The number of sets with one
element depends on the payoff matrix and the number of players
participating the game. Therefore, based on the above
observations, especially ({\it O5}), we can immediately conclude
that for multiparty extensions of $2\times 2$ games classified
into {\it Group II}, there will always be a contradiction for the
states $|W_N\rangle$ and $\ket{N-m,m}/\sqrt{{}_{N}C_m}$ except for
$\ket{1,1}/\sqrt{2}$ and $\ket{2,2}/\sqrt{6}$. Hence, the results
obtained for SCR are valid for WCR as well.

\subsubsection{Multiplayer games according to WCR} In the previous
subsection, we showed that the results of SCR are valid in case of
WCR for multiparty extensions of two-player two-strategy games.
Here, we consider the class of games which are originally designed
as multiplayer games:

{\it For the Minority game,} the payoff structure is such that
there is no set with odd number of elements therefore we cannot
exploit ({\it O5}). However, we have ({\it O1}) which is valid for
$|W_N\rangle$ and Dicke states $\ket{N-m,m}$ with $N\neq m$.  For
the Dicke states with $N=m$, the situation mentioned in ({\it O4})
occurs only for the state $\ket{2,2}$ because pairs of output
states leading to relations as in Eq. \ref{N13} are in the same
sets. Hence, for this state there will be {\it no contradiction}.
On the other hand, when $N\geq6$ the type of {\it contradictions}
described in ({\it b1})-({\it b3}) are seen. Hence, the results
obtained for the case of SCR are valid for Minority game. Consider
$N=4$ for which the payoff structure imposes the partitions
$S_1=\{\phi_{1,4,6,7,10,11,13,16}\}$, $S_2=\{\phi_{2,15}\}$,
$S_3=\{\phi_{3,14}\}$, $S_4=\{\phi_{5,12}\}$ and
$S_5=\{\phi_{8,9}\}$. The outcomes differing with the strategies
of four players are in the same sets. Therefore, for the Dicke
state with $N=m=2$ there will be no contradiction and this state
can be used. For $|W_4\rangle$, mutual orthogonality of set-pairs
$(S_1,S_{3,4,5})$ requires $\langle \phi_1|\phi_{3,5,9}\rangle=0$
which gives $\hat{u}_k^{1\dagger}\hat{u}_k^2 =\hat{\sigma}_x
\hat{R}_z(2\phi_k)$ for $\forall k$.  Substituting in the
orthogonality relations from $(S_5,S_{2,3,4})$ we obtain $\langle
\phi_{2,3,5}|\phi_8\rangle=0$ which gives $\chi_{234}$ implying a
{\it contradiction}.

In a {\it coordination game}, the players receive the payoffs
$\lambda_0>0$ ($\lambda_1>0$), when all choose the first (second)
strategy; otherwise, they receive zero. If
$\lambda_0\neq\lambda_1$, players make their choices for the
strategy with the higher payoff. A game-theoretic situation occurs
only when $\lambda_0=\lambda_1$, because players cannot coordinate
their moves without communication. The payoff structure and
outcomes of such a game can be grouped into two sets; the first
one will have two elements, where all players choose either the
first or second strategy, $S_1=\{\phi_1,\phi_{2^N}\}$, and the
second one, $S_2$ will have the rest of the outcomes. In such a
partition all the contradictions mentioned above will appear
except for the state $|2,2\rangle$. If $\lambda_0\neq\lambda_1$
then we will have three sets two of which will be with one
element, and hence the observation ({\it O5}) will be valid.
Therefore, we conclude for this game and any other game with such
a payoff structure, all the results of SCR are valid.

For a {\it majority game}, all the players receive $\lambda_0$ or
$\lambda_1$ depending on whether the majority is achieved in the
first or second strategy, respectively. In case of even-split all
get zero. Outcomes are grouped into three and four sets for odd
and even $N$, respectively. For both cases all the results
obtained for SCR is valid. Here we give the examples for $N=3$ and
$N=4$. When $N=3$, outcomes are grouped as
$S_1=\{\phi_{1,2,3,5}\}$ and $S_2=\{\phi_{4,6,7,8}\}$. From
$\langle\phi_{2,3}|\phi_4\rangle$ and
$\langle\phi_{2}|\phi_6\rangle$ we obtain $\hat{u}_k^{1\dagger}
\hat{u}_k^2=\hat{\sigma}_x \hat{R}_z(2\phi_k)$ for $\forall k$.
Then $\langle\phi_{1}|\phi_{4,6}\rangle$ and
$\langle\phi_{2}|\phi_{8}\rangle$ results in $\chi_{123}$. This is
exactly the situation in ({\it O1}). For $N=4$, the outcomes are
divided into three sets as $S_1=\{\phi_{1,2,3,5,9}\}$,
$S_2=\{\phi_{8,12,14,15,16}\}$ and $S_3=\{\phi_{4,6,7,10,11,13}\}$
where $S_3$ has the outcomes for the even-split of choices. For
$|W_4\rangle$ and $\ket{N-m,m}$ with $N\neq m$, we have
$\hat{u}_k^{1\dagger} \hat{u}_k^2=\hat{\sigma}_x
\hat{R}_z(2\phi_k)$ for $\forall k$ from
$\langle\phi_{2}|\phi_{4,6,10}\rangle=0$ and
$\langle\phi_{3}|\phi_{4}\rangle=0$ due to the orthogonality of
$(S_1,S_3)$. Moreover, we have
$\langle\phi_{1}|\phi_{4,6,7}\rangle=0$  which results in the set
$\chi_{234}$. This is also exactly the situation in ({\it O1}).
Similar contradiction can be obtained from the orthogonality of
$(S2,S3)$, too. On the other hand when we use $|2,2\rangle$, one
can show that no contradictions occur and the strategies can be
chosen as $\hat{u}_k^{1}=\hat{\sigma}_0$,
$\hat{u}_1^{2}=\hat{\sigma}_x$ and
$\hat{u}_2^{2}=\hat{u}_3^{2}=\hat{u}_4^{2}=(\sqrt{2}\hat{\sigma}_z+\hat{\sigma}_y)/\sqrt{3}$.

In a {\it zero-sum game} where there is competitive advantage
$\lambda$, if all players choose the same strategy, there is no
winner and loser so all receive zero. Otherwise, each of the $m$
players choosing the first strategy gets $\lambda/m$, and the rest
of the players loses $\lambda/(N-m)$. The outcomes are grouped
into $2^N-1$ sets where one set has two elements obtained when all
players choose the same strategy and the rest with one element. In
this case, ({\it O6}) is valid and hence there will be {\it
contradiction}. In the multi-player extension of MP game, the
outcomes when all players choose the same strategy are always
grouped into one set of two elements, and the rest of the outcomes
are grouped in sets of even-number of elements (the number of sets
depends on $N$). Thus, ({\it O6}) is observed, and hence the same
results are valid.

A {\it symmetric game} with a strict ordering of the payoffs is a
{\it Group I}; otherwise  a {\it Group II} game. We analyzed such
games up to $N=6$, and found that all the results concerning the
entangled states and operators are valid except for a few
exceptional cases which we could not relate to any game-theoretic
situation when $N=3$ and $N=6$. For $N=3$, we have eight outcomes
with the payoff vectors as $(a,a,a)$, $(b,b,d)$, $(b,d,b)$,
$(c,e,e)$, $(d,b,b)$, $(e,c,e)$, $(e,e,c)$ and $(f,f,f)$. With
proper choices of the parameters, one can obtain multiplayer
extensions of the symmetric games, PD, MD, RC, CG, SH and AG. For
other possible generic games, we search for the values of the
payoff entries for which there will be no contradiction according
to discussions above. For the entangled states $|W_3\rangle$ and
$\ket{3-m,m}/\sqrt{{}_{3}C_m}$, we know from ({\it O5}) that all
sets must have even number of elements. We identify five possible
partitions (2 two-set partitions and 1 one-, three- and four- set
partitions): (1) One- and four-set partitions require all outcomes
be the same, $a=b=c=d=e=f$, that is all players receive the same
payoff no matter which strategy they choose. This is not a game.
(2) Three-set partitions result in three different cases:  (i)
$a=b=d$ and $c=e=f$ which is the majority game discussed above,
(ii) $a=b=c=d=e=f$ as in (1), and (iii) $a=c=e=\lambda_0$ and
$b=d=f=\lambda_1$ where payoffs of the players are equal
regardless of their choice. Players receive $\lambda_0$ when
two-players choose the second strategy and one chooses the first
or when they all choose the first strategy; otherwise they receive
$\lambda_1$. Such a situation does not correspond to a
game-theoretic one. (3) Two-set partitions, in addition to those
listed in (2), result in $a=f$ and $b=d=c=e$ which corresponds to
coordination game discussed above. In the case of the Dicke state
For $\ket{N/2,N/2}$ with $N=6$, no contradiction occurs if the
outputs are divided into two sets each with thirty-two elements.
The first set includes the outcomes when four players choose the
first strategy and two choose the second strategy, when all
players choose the first strategy,and when all choose the second
strategy. The rest of the outputs are in the second set. We could
not find any game-theoretic situation with such a payoff
structure. Thus, the results obtained so far are valid for up to
six-player symmetric games which represent a game-theoretic
situation and hence are the subject of game theory.

\subsection{Reproducibility criterion as a benchmark} It is only
when reproducibility criterion is satisfied, we can compare the
outcomes of classical and quantum versions to draw conclusions on
whether one has advantage over the other. The first thing the
physical scheme should provide is unitary operators corresponding
to classical pure strategies for a given entangled state. If there
exists such operators then one can compare the outcomes for the
pure strategies. Let us consider the entangled state $|W_N\rangle$
for which one cannot find $\{\hat{u}_k^{1},\hat{u}_k^{2}\}$
satisfying the criterion. When a game is played using
$|W_N\rangle$ with unitary operators chosen from the SU(2), the
outcomes of the classical game in pure strategies cannot be
obtained, because in the quantum pure strategy, the payoffs become
a probability distribution over the entries of the classical
payoff matrix. Therefore, comparing the quantum version using
$|W_N\rangle$ with the classical game in pure strategies is not
fair. In the same way, comparing quantum versions played with GHZ
and $|W_N\rangle$ states is not fair either because for GHZ the
payoffs delivered to the players are unique entries from the
classical payoff table, contrary to those for $|W_N\rangle$. Thus,
we think the reproducibility criterion constitutes a benchmark not
only for the evaluation of entangled states and operators in
quantum games but also for the comparison of classical games and
their quantum versions.

\section{Conclusion} In this paper, for the first time, we
give the necessary and sufficient  condition to play quantized
version of classical games in a physical scheme. This condition is
introduced here as the {\it reproducibility criterion} and it
provides a fair basis to compare quantum versions of games with
their classical counterparts. This benchmark requires the
reproducibility of the results of the classical games in their
quantum version. The SCR and WCR shows that a large class of
multipartite entangled states cannot be used in the quantum
version of classical games; and the operators that might be used
should have a special diagonalized form. Given two unitary
operators $\{\hat{u}_k^{1},\hat{u}_k^{2}\}$ corresponding to
classical pure strategies and satisfying SCR and/or WCR, one can
reproduce the results of classical games in pure strategies in the
physical scheme. Moreover, provided that the players choose
unitary operators in the space spanned by $\hat{u}_k^{1}$ and
$\hat{u}_k^{2}$, mixed strategy results of classical games can be
reproduced, too. The results are valid for a large class of
entangled states, which can be prepared experimentally with the
current level of technology, and multi-player extensions of
interesting $2\times 2$ games as well as for a large class of
originally multiparty games.

Results also suggest that entangled states that cannot be used in
two-strategy multi-player games due to SCR are good candidates for
quantum information tasks (i.e, multi-party binary decision
problems, etc) where anonymity of participants is required. SCR
can be rephrased as the construction of complete orthogonal bases
from an initially entangled state by local unitary operations when
the parties are restricted to a limited number of operators. While
this construction is possible for the states satisfying SCR, it is
not possible for the others.

Extending this work to any generic game and the whole family of
$N$-partite entangled states requires lengthy calculations and
detailed classification of payoff structures which is beyond the
scope of this paper. However, the results presented here are
enough to show the importance of reproducibility criterion and the
restrictions imposed by it. \ack Authors thank M. Koashi, F.
Morikoshi and T. Yamamoto for their support and useful
discussions.

\end{document}